\documentstyle[12pt]{article}
\begin{document}


\renewcommand{\thesection}{\arabic{section}}
\renewcommand{\theequation}{\arabic{equation}}
\renewcommand {\c}  {\'{c}}
\newcommand {\cc} {\v{c}}
\newcommand {\s}  {\v{s}}
\newcommand {\CC} {\v{C}}
\newcommand {\C}  {\'{C}}
\newcommand {\Z}  {\v{Z}}

\baselineskip=24pt


\begin{center}
{\bf   Matrix oscillator and Calogero-type models}
 
\bigskip

S.Meljanac {\footnote{e-mail: meljanac@irb.hr}} \hspace{0.3cm} and \hspace{0.3cm}  
A. Samsarov{ \footnote{e-mail:asamsarov@irb.hr}}\\
 Rudjer Bo\v{s}kovi\'c Institute, Bijeni\v cka  c.54, HR-10002 Zagreb,
Croatia\\[3mm]

\bigskip

\end{center}
\setcounter{page}{1}
\bigskip


\begin{center}
{\bf   Abstract 

\bigskip}

\end{center}

 We study a single matrix oscillator with the quadratic Hamiltonian and deformed commutation relations. It is equivalent
to the multispecies Calogero model in one dimension, with inverse-square two-body and three-body interactions.  
Specially, we have constructed a new matrix realization of the
Calogero model for identical particles, without using exchange operators. The critical points at which singular
behaviour occurs are briefly discussed.

\bigskip
PACS number(s): 03.65.Fd, 03.65.Sq, 05.30.Pr 


\newpage




\section {Introduction}
 A class of integrable many-body systems in one spatial dimension is known, referred to as  Calogero systems [1].
These systems are formed of $ N $ identical particles on the line which interact through an inverse-square two-body 
interaction and are subjected to a common confining harmonic force. These models are completely integrable 
in both the classical and the quantum case [2] and are related to a number of mathematical and physical problems,
ranging from random matrices [3,4] to gravity, black hole physics [5]
and two-dimensional strings [6].
The algebraic structure of the Calogero model has recently been reconsidered by a number of authors in the framework of 
the exchange operator formalism [7,8] based on a symmetric group.
 An advantage of this approach is the 
possibility of an explicit construction of wave functions for an arbitrary number of particles. This approach also 
emphasizes the interpretation in terms of generalized statistics [9] that allows for the possibility of
 having particles of different species with a mutual coupling parameter depending on the species coupled.

  It is known that a random matrix theory provides a simple relation between the quantum mechanics of the harmonic
oscillator and the Calogero model. However, this connection is known to be true only for three special values of the
coupling parameter $ \nu $ : $ \nu = \frac{1}{2}, 1, 2 $. In this approach the Calogero model appears through the 
calculation of averages in the Gaussian ensembles [3,4]. Also, another remarkable connection between
the matrix models and the Calogero models was established in [2,10,11,12]. In Ref's.[2,11,12] a classical 
matrix system without the quadratic potential was considered by the technique of the Hamiltonian reduction.
The quantization is performed through  path integral methods [12].
\newpage

 In the present Letter we introduce matrices 
whose matrix elements are operators and define the matrix Hamiltonian of the quadratic type.
We show that this matrix formulation is in one-to-one correspondence with the Calogero model for an
arbitrary value of the coupling parameter that can even depend on the particles coupled. In addition, exchange operators
do not appear in our formulation.

In section $ 2 $, by introducing a pair of $ N \times N $ matrices   $ \bf{X}$  ,  $ \bf{P}$, we define the 
quadratic Hamiltonian $ \mathcal{H} $ for a single quantum matrix oscillator that is required to satisfy
deformed commutation relations. This is a generalization of a single oscillator with the quadratic Hamiltonian
and the deformed commutation relation [13,14].
After finding the representation that solves these commutation relations, we show that the single matrix
oscillator with the quadratic Hamiltonian and  deformed commutation
relations is equivalent to a multispecies Calogero model [15-18] with inverse-square two-body and three-body interactions.
 Generalization of $ SU(1,1) $ generators is proposed and their form 
is used to construct matrix ladder operators. By applying the Fock space analysis, a class of the excited states 
of the matrix oscillator Hamiltonian $ \mathcal{H} $ has been found. In section $ 3 $  we specialize our 
considerations to the case where all masses and coupling constants are equal. After stating some non-trivial identities, 
the matrix oscillator considered in this paper turns out to be an alternative simple formulation of the Calogero
model that avoids the necessity of using exchange operators. Finally, brief inspection of the
Fock space that corresponds
to the relative motion of particles reveals the existence of the critical points at which the system exhibits 
singular behaviour.



\section{ Matrix oscillator and a multispecies Calogero model}
  Let us consider  $ N \times N $ matrices   $ \bf{X}$  ,  $ \bf{P}$  with operator-valued matrix elements and 
a non-singular mass matrix  $ \; \mathcal{ M}$. The matrix Hamiltonian, generally non-Hermitean, is given by 
\begin{equation}
  {\mathcal{H}} =  \frac{1}{2} ( {\bf{P}}  {\mathcal{ M}}^{-1} {\bf{P}} + {{\omega}^{2}}
 {\bf{X}}  {\mathcal{ M}}  {\bf{X}} ),
 \end{equation}
with   $ \hbar = 1 $. It represents a matrix generalization of a single harmonic oscillator.\\
We assume the following matrix commutation relations:
\begin{equation}
  [ \bf {X} , \bf {P}] = \imath \mathcal{V},
\end{equation}
where  $  \mathcal{V} $  is a Hermitean  $ N \times N $  matrix with  constant, real and symmetric 
 matrix elements  $ \nu_{ij} = \nu_{ji} , \;\;  i,j = 1,2,...,N $. We set the diagonal elements
to be equal to unity, reflecting the quantum nature of the system. We further assume that the
matrix $\bf{X}$ is Hermitean and can be represented as a diagonal matrix with real elements
 $ x_{i} , \;\; i = 1,2,...,N $ :
\begin{equation}
 {\bf{X}}_{ij} = x_{i}\delta_{ij}
\end{equation}
and  $ {\mathcal{ M}}_{ij} = m_{i}\delta_{ij}, \;\; m_{i} > 0 $ .
 By introducing the matrix operator $ \mathcal{ D}$ such that the relation $ \; \bf{ P} = -\imath \mathcal{ D} \; $ holds,
we can rewrite Eq. (2) in the  following way :
\begin{equation}\begin{array}{c}
 {\mathcal{D}}_{ij} x_{j} - x_{i}{\mathcal{D}}_{ij} = \nu_{ij} \;\;\;\;  \forall i,j ; \;\; i \neq j  \\
 {\mathcal{D}}_{ii} x_{i} - x_{i}{\mathcal{D}}_{ii} = 1 \;\;\;\;   \forall i .
\end{array}\end{equation}
There are many solutions of equations (4) since the addition of a diagonal piece to $ {\mathcal{D}}_{ij} $ depending
 only on the coordinates does not affect these equations. To express this fact in a more explicit maneer, we can
 consider a transformation $ \; {\mathcal{ D}}^{(f)} = f^{-1} {\mathcal{ D}} f \; $ of the operator $ \; \mathcal{ D} \; $
 by the arbitrary function $ \; f \; $ of the coordinates. The corresponding Hamiltonians are connected by non-unitary
 gauge transformations, i.e. by  similarity transformations
of the form $ \; {\mathcal{ H}} ^{(f)} = f^{-1} {\mathcal{ H}} f \; $.
We shall restrict ourselves to gauge transformations defined by 
$ \; f = \prod_{i < j}(x_{i} - x_{j})^{\lambda_{ij}}, \;\;\; \lambda_{ij} = \lambda_{ji} $.
 A corresponding class of solutions of equations (4) is given by
\begin{equation}
 {\mathcal{D}}_{ij} = \delta_{ij} ( \frac{\partial}{\partial x_{i} } + \sum_{k \neq i}\frac{\lambda_{ik}}{x_{i} - x_{k}} )
   -  \frac{\nu_{ij}(1 - \delta_{ij})}{x_{i} - x_{j}},
\end{equation}
where $ \; \lambda_{ij}  \; $ are gauge parameters.
Note that the dependence of the Hamiltonian $ \;  \mathcal{ H} \; $ on the 
gauge parameters $ \;  \lambda_{ij} \; $ enters through the operator  $ \mathcal{ D}$.

The matrix Schroedinger equation is
\begin{equation}
 {\mathcal{H}} \ast   \Psi ({\bf{X}})  \equiv  {\mathcal{H} \mathcal{J}}  \Psi({\bf{X}})\;
   = {\mathcal{E}}   \Psi({\bf{X}}),
 \end{equation}
where $ \Psi(\bf{X})$ is a column wave function $ ( \psi_{i}(\bf{X})), \;\;$  $ i = 1,...,N $, 
 $  \mathcal{J} $ is an $ N \times N $ matrix with units at all positions,
and multiplication $ \ast $ is defined in the above equation. For example, the ground state in the gauge 
 $ \; \lambda_{ij} = \nu_{ij}, \;\;\; \forall  i,j \; $ is described by the column matrix
\begin{equation}
 \| 0  \rangle \sim  \left ( e^{-\frac{\omega}{2}\sum_{i=1}^{N} m_{i}x_{i}^{2}}\right ) \; C .
\end{equation}
Here  $ C $ is the column matrix with all elements equal to unity.
Analogously, one can introduce the left action of the Hamiltonian $ \mathcal{H}$
 on the row wave function. In this case (in the gauge  $ \; \lambda_{ij} = \nu_{ij}, \;\;\; \forall  i,j \; $), the 
ground state is represented by $ \; \langle 0 \| \sim  R e^{-\frac{\omega}{2}\sum_{i=1}^{N} m_{i}x_{i}^{2}} \; $,
 where  $ R $ is a transpose of  $\; C $.
 Note that  $ \;  RC = N  $,  $ \;  CR =  {\mathcal{J}}, \;  {\mathcal{J}}\| 0  \rangle = N \| 0  \rangle \; $
  and  $ \;  \langle 0 \|0 \rangle = 1 $.
 At this point it should be emphasized that  $ \;\; \Psi ({\bf{X}})  \;\; $ is not the eigenstate of the Hamiltonian
 $ \; {\mathcal{H}} \; $ in the usual sense, but rather it satisfies the ``eigenvalue'' equation in which the 
 ``eigenvalue''  $ \; {\mathcal{E}}\; $ is a matrix. We can give a well-defined meaning to this equation
  after performing multiplication of both sides by the row matrix $ \; R  \; $ from the left.
 In this case, Eq. (6) is reduced to the  eigenvalue
equation $ \; H\psi = E \psi \;  $, where $ \; H \; $ is the Hamiltonian corresponding to the matrix Hamiltonian 
$ \; {\mathcal{H}} \; $ by  $ \; H = R {\mathcal{H}} C =  tr( {\mathcal{H}} {\mathcal{J}}) \; $ .
 We point out that in the special case where the gauge 
$ \; \lambda_{ij} = \nu_{ij}, \;\;\; \forall  i,j \; $ is chosen, we get the familiar eigenvalue equation
$ \; \tilde{H} \psi = E \psi, \;  $ where  $ \; \tilde{H} \; $ stands for the transformed Hamiltonian for a multispecies
 Calogero model [17]. In this case   $ \; {\mathcal{H}} \;$ is related to $ \; \tilde{H} \; $ as
$$
  \tilde{H} = R {\mathcal{H}} C = tr (\mathcal{H} \mathcal{J}) 
$$
$$
  = -\frac{1}{2}\sum_{i=1}^{N}\frac{1}{m_{i}}\frac{{\partial}^{2}}
{\partial x_{i}^{2}} + \frac{{\omega}^{2}}{2}\sum_{i=1}^{N }m_{i} x_{i}^{2}
$$
$$
- \frac{1}{2} \sum_{i \neq j }\frac{{\nu}_{ij}}
{(x_{i}-x_{j})}(\frac{1}{m_{i}} \frac{\partial}{{\partial} x_{i}}
 - \frac{1}{m_{j}} \frac{\partial}{{\partial} x_{j}})
$$
\begin{equation}
 \equiv - T_{-} + {{\omega}^{2}} T_{+},
\end{equation}
where $ \; T_{\pm} \; $ are $ \; SU(1,1) \; $ generators. 
After performing the similarity transformation with the inverse Yastrow 
factor $ \; \prod_{i < j}(x_{i} - x_{j})^{ - {\nu}_{ij}} \; $, we obtain an original Hamiltonian
$ H_{cal} $ for the multispecies Calogero model,
with inverse-square two-body and three-body interactions [17]. This original Hamiltonian $ H_{cal} $ can also be
 reproduced directly, $ H_{cal} = tr( {\mathcal{H}} {\mathcal{J}}) $  in the 
 gauge $ \; \lambda_{ij} = 0, \;\;\; \forall  i,j  $.
The procedure outlined here differs significantly from that followed in  Ref.[12] in the way how the
 Calogero Hamiltonian appears.
Namely, in this paper the Hamiltonian is not just the trace of the matrix Hamiltonian  $ \mathcal{H}, $ but is given by
$ \; H = tr( {\mathcal{H}} {\mathcal{J}}) , \; $ see Eq.(8). In the rest of the paper we restrict to the gauge 
$ \; \lambda_{ij} = \nu_{ij}, \;\;\; \forall  i,j \; $. 

In the same way as we have introduced the matrix Hamiltonian $ \mathcal{H} $, we introduce matrix generators with operator-
 valued matrix elements
\begin{equation}\begin{array}{l}
 {\mathcal{T}}_{+} =  \frac{1}{2} \bf{X} \mathcal{ M} \bf{X},  \\
 {\mathcal{T}}_{-} =  \frac{1}{2} {\mathcal{D}} {\mathcal{ M}}^{-1} {\mathcal{D}}, \\
 {\mathcal{T}}_{0} =  \frac{1}{4}( \bf{X}{\mathcal{ D}} + {\mathcal{ D}}\bf{X}) 
                        = \frac{1}{2} ( \bf{X}{\mathcal{ D}} + \frac{1}{2} {\mathcal{ V}}). 
\end{array}\end{equation}
They satisfy the relations
\begin{equation}
  R [{\mathcal{T}}_{-}, {\mathcal{T}}_{+}]_{J}C = 2 R {\mathcal{T}}_{0}C ,
\end{equation}
\begin{equation}
 R [{\mathcal{T}}_{0}, {\mathcal{T}}_{\pm}]_{J}C = \pm R {\mathcal{T}}_{\pm}C ,
\end{equation}
where the  $ J - \mbox{commutator} $ is defined by 
\begin{equation}
  [A,B ]_{J} = A {\mathcal{J}} B - B {\mathcal{J}} A .
\end{equation}
$ {\mathcal{T}}_{\pm}, {\mathcal{T}}_{0} \;\; $  are related to the generators $ \;\; T_{\pm}, T_{0}   \;\;$ [17] of
$ \; SU(1,1) \;$ algebra in the following way :
\begin{equation}\begin{array}{l}
 T_{\pm} = R {\mathcal{T}}_{\pm} C = tr ({\mathcal{T}}_{\pm}{\mathcal{J}}), \\ 
 T_{0} = R {\mathcal{T}}_{0} C = tr ({\mathcal{T}}_{0}{\mathcal{J}}) . 
\end{array}\end{equation}

The wave functions $ \psi $ of the
Hamiltonian (8) $ ( \tilde{H}\psi = E \psi ) $ are related to the  column wave functions defined in (6). As it is readily
seen, the connection is simply $ \psi ( x_{1},..., x_{N}) \sim  R \Psi(\bf{X}) \; $ and
$ \; {\psi ^{\ast}}  ( x_{1},..., x_{N}) \sim {\Psi}^{\dagger} ({\bf{X}}) C \;  $. 

 The model described by the Hamiltonian (8) was partially solved in [17]. Corresponding solutions in the
matrix formulation are  obtained in terms of the following  pairs of creation and annihilation operators :
$$
 {\mathcal{A}}_{1}^{\pm} = \frac{1}{\sqrt{2 tr {\mathcal{M}}}}( \sqrt{\omega}{\bf {X}}{\mathcal{M}} \mp 
     \frac{1}{\sqrt{\omega}}{\mathcal{D}} ),
$$
\begin{equation}
  {\mathcal{A}}_{2}^{\pm} = \frac{1}{2}
 ( \omega {\mathcal{T}}_{+} + \frac{1}{\omega} {\mathcal{T}}_{-}) \mp {\mathcal{T}}_{0} .
\end{equation}
Note that for the case in which all masses $\; m_{i}\;$ are equal, there is a simple relation between these two sets
of operators, namely $ \;\; N R {{\mathcal{A}}_{1}^{\pm}}^{2}C = R {\mathcal{A}}_{2}^{\pm}C   \;\;$ .

  The generators $ \;  {\mathcal{T}}_{\pm} \; $ are  defined in (9) and  play the role of collective radial variables
corresponding to dilatation modes. The first pair of operators (14) describes center-of-mass ($ CM $) modes,
 while the second pair describes collective radial modes. These operators satisfy the following commutation relations :
$$
 R [ {\mathcal{A}}_{1}^{-}, {\mathcal{A}}_{1}^{+}]_{J}C = 1 ,  \;\;\;\;\;
 R [ {\mathcal{A}}_{2}^{-}, {\mathcal{A}}_{2}^{+}]_{J}C = \frac{1}{\omega} R {\mathcal{H}}C,
$$
$$
 R [ {\mathcal{A}}_{1}^{\mp}, {\mathcal{A}}_{2}^{\mp}]_{J}C = 0, \;\;\;\;\;
 R [ {\mathcal{A}}_{1}^{\mp}, {\mathcal{A}}_{2}^{\pm}]_{J}C = \pm R {\mathcal{A}}_{1}^{\pm}C , 
$$
\begin{equation}
 R [ {\mathcal{H}}, {\mathcal{A}}_{1}^{\pm}]_{J}C = \pm \omega R {\mathcal{A}}_{1}^{\pm}C  , \;\;\;\;\;
 R [ {\mathcal{H}}, {\mathcal{A}}_{2}^{\pm}]_{J}C = \pm 2 \omega R {\mathcal{A}}_{2}^{\pm}C .
 \end{equation}  

 The partial matrix Fock space corresponding to  $ CM $ modes and collective 
radial modes is spanned by the states of the form 
\begin{equation}
 \| n_{1},n_{2} \rangle \sim ({\mathcal{A}}_{1}^{+} {\mathcal{J}})^{n_{1}}({\mathcal{A}}_{2}^{+} {\mathcal{J}})^{n_{2}}
   \| 0  \rangle,
\end{equation}
which are governed by the matrix Schroedinger equation
\begin{equation}
   {\mathcal{H}}{\mathcal{J}} \| n_{1},n_{2} \rangle =  {\mathcal{E}}_{n_{1},n_{2}}\| n_{1},n_{2} \rangle .
\end{equation}
Here $ \;\; {\mathcal{E}}_{n_{1},n_{2}} \;\; $ is the matrix that satisfies the relation 
$$
  \frac{1}{\sqrt{N}} R  {\mathcal{E}}_{n_{1},n_{2}}\| n_{1},n_{2} \rangle = 
  \frac{1}{\sqrt{N}} E_{n_{1},n_{2}} R \| n_{1},n_{2} \rangle 
$$
\begin{equation}
 = E_{n_{1},n_{2}} | n_{1},n_{2} \rangle = ( E_{0} +  \omega (n_{1} + 2n_{2})) | n_{1},n_{2} \rangle ,
\end{equation}
where $ \;  E_{0} = \omega ( \frac{N}{2} + \frac{1}{2}\sum_{i \neq j} \nu_{ij})\; $ is the energy of the
 ground state. The state 
$ \;  \| n_{1},n_{2} \rangle \; $ is the column Fock  state and $ \;  \| 0  \rangle \; $ is the vacuum state
defined by (7), while
$ \;\;  E_{n_{1},n_{2}}   \;\; $  and  $ \;\; | n_{1},n_{2} \rangle  \;\; $ are eigenvalues and eigenstates
of the partial Fock space of the corresponding multispecies Calogero problem.  
 The ground state is well defined if $ \;  E_{0} > \frac{1}{2} \; $ [17-19].

Note that the correspondence between the matrix ladder operators (14) and the analogous operators
  $ \;A_{1}^{\pm},   A_{2}^{\pm} \; $
  [17] that define the partial Fock space in the related multispecies Calogero model is simply given by
$$
 A_{1}^{\pm} = R {\mathcal{A}}_{1}^{\pm} C = tr ( {\mathcal{A}}_{1}^{\pm}{\mathcal{J}} ),
$$ 
\begin{equation}
 A_{2}^{\pm} = 
   R {\mathcal{A}}_{2}^{\pm} C = tr ( {\mathcal{A}}_{2}^{\pm}{\mathcal{J}} ).
\end{equation}

\section{ Special case : the Calogero model}

Specially, for $ \; \nu_{ij} = \nu \; $,  $ \; m_{i} = m $  for all $ \;  i,j = 1,...,N \; $, we recover the original 
Calogero model ( cf. Eq. (8)).  

In this case, we introduce the matrix creation and annihilation operators $ {\mathcal{A}}^{\pm} $ defined by
\begin{equation}
 {\mathcal{A}}^{\pm} = \frac{1}{\sqrt{2 m \omega}} ( m \omega {\bf{X}} \mp  {\mathcal{D}}) =
       \frac{1}{\sqrt{2 m \omega}} (m \omega {\bf{X}} \mp \imath \bf{ P} ).
\end{equation}

Then the following relations hold : 
\begin{equation}
 [{\mathcal{A}}^{-}, {\mathcal{A}}^{+}] = (1 - \nu){\bf {1}} + \nu \mathcal{J}              
\end{equation}
and
\begin{equation}
 {\mathcal{H}} = \frac{\omega}{2} \{{\mathcal{A}}^{-}, {\mathcal{A}}^{+}\}, \;\;\;\;
 R {\mathcal{H}}C   = \omega R [{{\mathcal{A}}^{-}}^{2} , {{\mathcal{A}}^{+}}^{2} ]_{J}C,              
\end{equation}
\begin{equation}
 R [{{\mathcal{A}}^{\mp}}^{k}, {{\mathcal{A}}^{\mp}}^{l} ]_{J}C = 0, \;\;\;\;\;
 R [{\mathcal{H}}, {{\mathcal{A}}^{\pm}}^{k} ]_{J}C = \pm \omega k R {{\mathcal{A}}^{\pm}}^{k}C ,
\end{equation}
where  $ \; k,l \; $ can be arbitrary non-negative integers.
In the above relations,
 $ [ , ] $ and $ \{ , \} $ denote the ordinary commutators and anticommutators, respectively.
Note that the set of operators  $ \{ {{\mathcal{A}}^{-}}^{2} ,{{\mathcal{A}}^{+}}^{2}  ,  {\mathcal{H}} \} $
define  $ \; SU(1,1) \; $ algebra and that the $ \; {\mathcal{A}} \; $ operators (20) coincide (up to the factor
 $ \; \sqrt{N}   $ )
 with the operators $ \; {\mathcal{A}}_{1}^{\pm} \; $ defined by (14),
for  $ \; m_{i} = m \;\; \mbox{and} \;\; \nu_{ij} = \nu  $,  namely
 $ \; {\mathcal{A}}^{\pm} = \sqrt{N} {\mathcal{A}}_{1}^{\pm}  \; $.

Starting from the column vacuum state (7) ( now with all $  m_{i} $ equal to $ m $ ), which satisfies the condition 
 $ \;\; {\mathcal{A}}^{-} \| 0  \rangle = 0 \;\; $, the complete matrix Fock space can be constructed.
 The general state of the full Fock space is generated by the operators (20) and  can be written as 
\begin{equation}
 \| n_{1},...,n_{N} \rangle \sim  \prod_{k=1}^{N}( {{\mathcal{A}}^{+}}^{k}{\mathcal{J}})^{n_{k}}
         \| 0  \rangle ,
\end{equation}
where  $ \;\; n_{k} = 0,1,2,... \;\; $ and
\begin{equation}
 {\mathcal{A}}^{-}{\mathcal{J}}{{\mathcal{A}}^{+}}^{n}\| 0  \rangle = 
  n {{\mathcal{A}}^{+}}^{n - 1}\| 0  \rangle .
\end{equation}

The general state, Eq(24), satisfies the equation
\begin{equation}
 {\mathcal{H}} \ast \| n_{1},...,n_{N} \rangle \equiv {\mathcal{H}}{\mathcal{J}}\| n_{1},...,n_{N} \rangle =
   {\mathcal{E}}_{\{n\}} \| n_{1},...,n_{N} \rangle .
\end{equation}
Here again, $ \;\; {\mathcal{E}}_{\{n\}} \;\; $ is the matrix that satisfies the relation
$$
 \frac{1}{\sqrt{N}}  R  {\mathcal{E}}_{\{n\}} \| n_{1},...,n_{N} \rangle = 
  \frac{1}{\sqrt{N}}  E_{\{n\}} R \| n_{1},...,n_{N} \rangle 
 $$
\begin{equation}
 = E_{\{n\}} | n_{1},...,n_{N} \rangle =
 ((\sum_{k=1}^{N} k n_{k} + \varepsilon_{0} ) \omega ) | n_{1},...,n_{N} \rangle ,
\end{equation}
where $ \;\; \varepsilon_{0} = \frac{1}{2}[ 1+ \nu (N - 1)] N \; $ and
$ \;\;  E_{\{n\}}   \;\; $  and  $ \;\; | n_{1},...,n_{N} \rangle  \;\; $ are eigenvalues and eigenstates
 of the corresponding  Calogero problem. For example, we explicitly show how the Hamiltonian
$\; {\mathcal{H}} \; $ acts on the state $ \;\; {{\mathcal{A}}^{+}}^{n}\| 0  \rangle \;\; $ :
\begin{equation}
     {\mathcal{H}} \ast {{\mathcal{A}}^{+}}^{n}\| 0  \rangle =
 \omega (n{\bf {1}} + \frac{1}{2}{\mathcal{V}} {\mathcal{J}}){{\mathcal{A}}^{+}}^{n}\| 0  \rangle .
\end{equation}
Note that after multiplying both sides of this equation by  $ R $ from the left, we get the familiar result.
Now we make  complete correspondence with the ordinary Calogero model for identical particles.
 
The Fock space for the ordinary single-species Calogero model is generated by totally symmetric combinations
 $ \;\; A_{k}^{\pm} = \sum_{i=1}^{N} ( a_{i}^{\pm} )^{k}   \;\;,(k = 1,2,...,N) \;\;  $
of auxiliary creation and annihilation  operators
 $ \;\; a_{i}^{\pm} = \frac{1}{\sqrt{2 \omega}}( \omega x_{i} \mp  \frac{\partial}{\partial x_{i}} \mp \nu \sum_{i \neq j}
   \frac{1}{x_{i}-x_{j}} ( 1 - K_{ij} )) \;\; $,  where  the operators $\;  K_{ij}\; $
  exchange labels $ \; i \; $ and $ \; j \; $ in all quantities [7,8].

To be more precise, the complete Fock space is spanned by the states of the form
 $ \;\;  \prod_{k=1}^{N} ( A_{k}^{+})^{n_{k}} |0 \rangle \;\;  $, where $ \;\; |0 \rangle \sim 
    e^{-\frac{ m \omega}{2}\sum_{i=1}^{N} x_{i}^{2}} \;\; $ is the ordinary vacuum, $  \;\; a_{i}^{-}|0 \rangle = 0 \;\;$
for all $ \;\; i = 1,...,N $.

At this point we emphasize that the following crucial relations hold:
\begin{equation}
 {(a_{i}^{+})}^{k}_{/_{symm}} = {( {{\mathcal{A}}^{+}}^{k} C )}_{i}, \;\;\;\;\;
 {(a_{i}^{-})}^{k}_{/_{symm}} = {( R {{\mathcal{A}}^{-}}^{k} )}_{i}
\end{equation}
for every $ \;\; k = 1,2,...,N \;\; $, which can be proved by induction.
 From this relation we immediately obtain
\begin{equation}
  A_{k}^{\pm} = \sum_{i=1}^{N}{(a_{i}^{\pm})}^{k}_{/_{symm}} =  R {{\mathcal{A}}^{\pm}}^{k} C =
   tr ( {{\mathcal{A}}^{\pm}}^{k} {\mathcal{J}}).
\end{equation}
The label  $ \; symm \; $ designates that the  action 
is restricted to the symmetric states only. In this case, the exchange operators
$ \; K_{ij} \; $ coincide with the identity operator $ ( K_{ij} \psi = \psi ) $.
Equations (29) provide  one-to-one correspondence between our matrix oscillator and the ordinary single-species 
Calogero model. Note that in the  matrix oscillator approach the exchange operators $ \; K_{ij} \; $ do not appear.
The approach considered in this paper can be repeated in an arbitrary gauge, particularly in the gauge $ \lambda = 0 $.

After removing the $ CM $ mode, one is left with $ \;\; {\mathcal{V}}_{rel} = ( 1 - \frac{1}{N} - \nu){\bf {1}}
 + \nu {\mathcal{J}}       \;\;$. The critical points are determined by  $ \;\; det {\mathcal{V}}_{rel} = 0 \;\; $.
They are $ \; \nu = - \frac{1}{N} \; $ and $ \; \nu = 1 -  \frac{1}{N} \; $. The first critical point corresponds to the 
existence of the ground state for $  \; \nu > -  \frac{1}{N} \; $ [17-19],
whereas the second critical point should be investigated more carefully.

\section{ Conclusion }
  In conclusion, by introducing a pair of  $ N \times N $ matrices, we have constructed a Hamiltonian for a single 
quadratic matrix oscillator that is required to obey  deformed commutation relations. This is a generalization of
a single deformed oscillator. Straightforward analysis has shown
that this oscillator is equivalent to the multispecies Calogero model in one dimension. Specially, in the case of
identical particles, an alternative formulation of ladder operators has been proposed in which exchange
operators do not appear.
In terms of these operators, the complete Fock space has been constructed.
We point out that our matrix formulation can be extended to all Calogero-like models, such as models in
arbitrary dimensions,the Sutherland model (on a circle) and models defined on root
systems and the supersymmetric Calogero model.
 
\bigskip
\bigskip



{\bf Acknowledgment}\\
We would like to thank K.S.Gupta and M.Milekovi\'{c } for useful discussions.
This work was supported by the Ministry of Science and Technology of the Republic of Croatia under 
contract No. 0098003.

\newpage




\end{document}